\input phyzzx
%
%
%
\newcount\lemnumber   \lemnumber=0
\newcount\thnumber   \thnumber=0
\newcount\conumber   \conumber=0

\def\myeq{{\rm \chapterlabel\the\equanumber}}

\def\Lemma{\par\noindent\global\advance\lemnumber by 1
           {\bf Lemma\ (\chapterlabel\the\lemnumber)}}
\def\Corollary{\par\noindent\global\advance\conumber by 1
           {\bf Corollary\ (\chapterlabel\the\conumber)}}
\def\Theorem{\par\noindent\global\advance\thnumber by 1
           {\bf Theorem\ (\chapterlabel\the\thnumber)}}

%
%
\def\e{\adveq\eqno{\rm (\chapterlabel\the\equanumber)}}
\def\adveq{\global\advance\equanumber by 1}
\def\twoline#1#2{\displaylines{\qquad#1\hfill(\adveq\myeq)\cr\hfill#2
\qquad\cr}}


%
%
\font\tensl=cmsl10
\font\tenss=cmssq8 scaled\magstep1
\outer\def\quote{
   \begingroup\bigskip\vfill
   \def\endquote{\endgroup\eject}
    \def\par{\ifhmode\/\endgraf\fi}\obeylines
    \tenrm \let\tt=\twelvett
    \baselineskip=10pt \interlinepenalty=1000
    \leftskip=0pt plus 60pc minus \parindent \parfillskip=0pt
     \let\rm=\tenss \let\sl=\tensl \everypar{\sl}}
\def\from#1(#2){\smallskip\noindent\rm--- #1\unskip\enspace(#2)\bigskip}

\def\WIS{\address{Department of Physics\break
      Weizmann Institute of Science\break
      Rehovot 76100, Israel}}

\def\r#1{$\lb \rm#1 \rb$}

%
%
\def\rarrow{\rightarrow}

\def\semidirect{\mathrel{\raise0.04cm\hbox{${\scriptscriptstyle |\!}$
\hskip-0.175cm}\times}}

\def\mod{\mathop{\rm mod}\nolimits}

\def\n{\mathopen{:}}
\def\nn{\mathclose{:}}
\def\ref#1{$^{#1}$}

\def\pr#1{{#1^\prime}}

\def\twidle{\tilde}

\def\half{{1\over2}}
\def\lb{\lbrack}
\def\rb{\rbrack}

\def\pr{\prime}
\def\diam{{\hbox{\hskip-0.02in
\raise-0.126in\hbox{$\displaystyle\bigvee$}\hskip-0.241in
\raise0.099in\hbox{ $\displaystyle{\bigwedge}$}}}}

%


\def\sqr#1#2{{\vcenter{\hrule height.#2pt
      \hbox{\vrule width.#2pt height#1pt \kern#1pt
        \vrule width.#2pt}
      \hrule height.#2pt}}}

\def\underwig#1{	
	\setbox0=\hbox{\rm \strut}
	\hbox to 0pt{$#1$\hss} \lower \ht0 \hbox{\rm \char'176}}

\def\bunderwig#1{	
	\setbox0=\hbox{\rm \strut}
	\hbox to 1.5pt{$#1$\hss} \lower 12.8pt
	\hbox{\seventeenrm \char'176}\hbox to 2pt{\hfil}}

\def\pr#1{{#1^\prime}}

\Pubnum={}
\pubtype={}
\date{January, 1999}
\titlepage
\title{On New Conformal Field Theories with Affine Fusion Rules}
\author{Doron Gepner}
\WIS
\abstract
Some time ago, conformal data with affine fusion rules were found. Our purpose
here is to realize some of these conformal data, using systems of free bosons
and parafermions. The so constructed theories have an extended $W$ algebras
which are close analogues of affine algebras. Exact character formulae is given,
and the realizations are shown to be full fledged unitary conformal field
theories.
\endpage
Rational conformal field theories have been the subject of much research owing
to their pivotal role in string theory and condensed matter physics, initiated
in the work of Belavin, Polyakov and Zamolodchikov \REF\BPZ{A.M. Polyakov, A.A. Belavin and A.B. Zamolodchikov, Nucl. Phys. B 241 (1984) 333.}
\r\BPZ. A relation later found by Verlinde \REF\Verlinde{E. Verlinde, Nucl. Phys. B 300 (1988) 360}
\r\Verlinde, connects the fusion rules and the modular matrix. In ref. \REF\Found{D. Gepner, Caltech preprint CALT-68-1825, hepth-9211100.}
\r\Found, it was noticed the the Verlinde formula has many,
as yet undescribed, solutions and it was conjectured that these are real
RCFT (rational conformal field theories). Our purpose here is to describe explicitly
the realizations of these data as full fledged rational conformal field
theories, for some families of subcases.

We concentrate on the affine-like conformal data, as the bosonic case was completely
solved in our recent publication \REF\Recent{
E. Baver, D. Gepner and U. Gursoy, preprint hepth-9811100} \r\Recent.
There we gave a formula for the central charge of the affine case. These
theories are labeled by an integer $q$, $G_{(k+g)/q}$ where $G$ is some group,
$k$ is the level and $q$ is any integer strange to $g(k+g)$. The theory
$G_{(k+g)/q}$ has the usual affine fusion rules \REF
\GW{D. Gepner and E. Witten, Nucl. Phys. B 278 (1986) 493.}\r\GW,
and the central charge,
$$c={qkD\over k+g} \mod 4,\e$$
where $g$ is the dual Coxeter number and $D$ is the dimension of the algebra.
$q=1$ is the usual affine case. Inspecting eq. (1), we notice that for
$q=p(k+g)+1$, where $p$ is some integer, the central charge obeys,
$$c(q)-c(1)={\rm integer}.\e$$
Thus, it is natural to suspect that for these cases $G_{(k+g)/q}$ is the same
as the usual affine case, up to some bosonic system. This is, in fact, our
result. More precisely, recall the parafermionic systems \REF\Para{V.A. Fateev and A.B. Zamolodchikov, Zh.Eksp.Theor.Fiz. 89 (1985) 215; D. Gepner and Z. Qiu, Nucl. Phys. B 285 (1987) 423; D. Gepner, Nucl. Phys. B 290 (1987) 10.}
\r\Para. The affine theories are decomposed into free bosons and parafermions,
whose characters are the so called string functions, $c^\Lambda_\lambda(\tau)$,
in such a way that
$$\chi^\Lambda(\tau)=\sum_\lambda c^\Lambda_\lambda(\tau) \Theta_\lambda(\tau),\e$$
where $\chi^\Lambda(\tau)$ is the affine character, and $\Theta_\lambda(\tau)$ are the
characters of a free bosonic system, propagating of the lattice $\sqrt{k}M$,
where $M$ is the root lattice. Mathematically, this formula was discussed
in \REF\KacBook{V.G. Kac, Infinite dimensional Lie algebras,
Cambridge Univ. press, 1990.}\r\KacBook, and ref. therein, and given a physical
interpretation in ref. \r\Para.

Our strategy, is to knock off the system of bosons in eq. (3) and to replace
it by another system of bosons, so as to get the theory $G_{(k+g)/q}$.
To this avail, we assume that the new bosonic system has the same fusion
rules as the old one. Such bosonic theories were realized fully in ref.
\r\Recent\ and are also labeled by some integer $\twidle q$, denoted by
$M_{k/\twidle q}$, where $M$ is the root lattice. The dimension formula for
the bosonic system is
$$\Delta_\lambda={\twidle q\lambda^2\over 2k}\mod Z.\e$$
In the pseudo-affine theories (i.e., $q\neq 1$), the fields have the
dimensions
$$\Delta_\Lambda={q\Lambda(\Lambda+2\rho)\over 2(k+g)}\mod Z,\e$$
where $\rho$ is half the sum of positive roots.

Combining, now eqs (4-5), we find the following relation for the dimensions,
$${(q-1)\Lambda(\Lambda+\rho)\over 2(k+g)}={(\twidle q-1)\lambda^2\over 2k}.\e$$
This relation can be fullfilled for $q-1=(k+g)p$ due to the relation
$\Lambda-\lambda\in M$. Taking $\twidle q=k\twidle p+1$, we find
$${p\Lambda(\Lambda+2\rho)\over2}={\twidle p \lambda^2\over 2}\mod Z.\e$$
This equation can easilly be satisfied, i.e., for all $p$ we can find
such $\twidle p$.

Examples: 1) Take $G\approx SU(2)$. Then, we take $\twidle p=3p$. Eq. (7)
becomes, where $\Lambda=l\alpha/2$ and $\lambda=m\alpha/2$,
$${p l(l+2)\over4}={\twidle p m^2\over 4}\mod Z,\e$$
which is obeyed for $\twidle p=3p\mod4$.

2) Take $G\approx SU(N)$ where $N$ is odd. Then we take $p=\twidle p$.
Eq. (7) becomes,
$${p \Lambda(\Lambda+2\rho)\over2}={\twidle p \lambda^2\over 2}\mod Z,\e$$
which is obeyed since $\rho\in M$ and $\Lambda-\lambda\in M$.
The same holds for all algebras with odd number of center elements.

3) Take $G=SU(N)$ where $N$ is even. It can be checked that it works exactly
like $SU(2)$ with $\twidle p=(N+1)p$. We omit the detail for
brevity sake.

For all the groups it is straightforward to solve the relation eq. (7).
Further, since we know all the realizations for such bosonic theories
\r\Recent, it follows that we have the complete realizations for all these
affine systems as full fledged unitary RCFT.

It remains, to verify the modular transformation $S:\tau
\rarrow -{1\over\tau}$. This is also quite simple since for the bosonic
systems we have the relation,
$$S_{\lambda,\mu}=\exp(-2\pi i[\Delta(\lambda+\mu)-\Delta(\lambda)-
\Delta(\mu)]),\e$$
we find that both sides of eq. (3) have the same modular transformations.
On the l.h.s. we have
$$S_{\Lambda,\pr\Lambda}=i^{|\Delta|}|{M^*\over (k+g)M}|^{-\half}\sum_{w\in W}
(-1)^w e^{-2\pi i q w(\Lambda+\rho)(\pr\Lambda+\rho)/(k+g)},\e$$
where $W$ is the Weyl group and $|\Delta|$ is the number of positive roots.
For $q=p(k+g)+1$ we get the phase $e^{-2\pi i
(\Lambda+\rho)(\pr\Lambda+\rho)/2}$ times the $q=1$ $S$ matrix, which is
exactly what is needed, by eq. (10).

Now, albeit ref. \r\Recent\ presents all the realizations for the bosonic
theories, the solution has to be worked out case by case. Further, we would
like to find the realizations with minimal central charge.
Using the methods of ref. \r\Recent\ it is not too difficult to present such
a construction.
This we do here for $SU(2)$.

First we introduce two types of matrices \r\Recent,
which depend on three integer parameters, $x,y,z$, obeying $x>0$, $y>0$ and $z^2<4 x y$. These matrices represent the scalar products between the basic vectors of the even lattices. 
The matrices are, of $E$ type,
$$E_{n}^{x,y,z} =\pmatrix{2 x &-z &0 &0  &\ldots&0&0&0&0\cr  
-z&2 y&-1 &0 &\ldots&0&0&0&0 \cr 
0&-1&2&-1&\ldots&0&0&0&0 \cr
0&0&-1&2&\ldots&0&0&0&0 \cr
\ldots&\ldots&\ldots&\ldots&\ldots&\ldots&\ldots&\ldots&\ldots& \cr
0&0&0&0&\ldots&2&-1&0&-1 \cr
0&0&0&0&\ldots&-1&2&-1&0 \cr
0&0&0&0&\ldots&0&-1&2&0 \cr
0&0&0&0&\ldots&-1&0&0&2},\e$$
for 
$$6 \le n \le 10,\qquad  \det(E_{n}^{x,y,z})=(4xy-z^{2})(11-n)-2x(12-n).\e$$

Of $A$ type,
$$A_n^{x,y,z}=\pmatrix{2 x &-z &0 &0  &\ldots&0&0&0&0\cr  -z&2 y&-1 &0 &\ldots&0&0&0&0 \cr 0&-1&2&-1&\ldots&0&0&0&0 \cr
0&0&-1&2&\ldots&0&0&0&0 \cr\ldots&\ldots&\ldots&\ldots&\ldots&\ldots&\ldots&\ldots&
\ldots& \cr
0&0&0&0&\ldots&2&-1&0&0 \cr
0&0&0&0&\ldots&-1&2&-1&0 \cr
0&0&0&0&\ldots&0&-1&2&-1 \cr0&0&0&0&\ldots&0&0&-1&2},\e)$$ 
$$ \det(A_n^{x,y,z})=(4 x y-z^2)(n-1)-2 x (n-2).\e$$

There are several cases. Introduce the usual theta function for the lattice $M$, 
$$\Theta_\lambda^M(\tau)=\sum_{\mu\in M+\lambda} e^{\pi i \tau\mu^2},\e$$
where $M$ is some lattice and $\lambda\in M^*\mod M$. First, we take for
$M$ the usual parafermionic lattice $M\approx (\sqrt{2k})$. However, we
permute the fields,
$$\chi_l=\sum_{m=l\mod2} c^l_m \Theta_{zm}.\e$$
we claim that for some $z$ this realizes the pseudo-affine theory,
for $k=0\mod8$.
From the formula for the dimensions, we need
$$z^2-1=\twidle p k,\e$$
which is solved as follows:

1) $k=0\mod 8$, $\twidle p=\pm1\mod4$: $\twidle 
p=k/4\mp 1$, $z=\mp1+k/2$.

2) $k=0\mod4$, $\twidle p=2\mod4$: $z=k\mp1$, $p=k\mp2$.

3) $k=0\mod4$, $\twidle p=1\mod4$: $M\approx 
E_9^{1,k/4+1,1}$ or $M\approx E_9^{k,k/4+1,k}$. The 
character is 
$$\chi^l=\sum_m c^l_m \Theta^M_{zm},\e$$
with
$z=1$. For $\twidle p=3\mod4$ it is the same lattice
with $z=k-1$ (for $k=0\mod8$), $z=k/2+1$ 
(for $k=4\mod16$), $z=k/2-1$ (for $k=12\mod16$).

4) For $k=2\mod4$, $p=1,2,3\ldots$. We have: $M\approx 
A_{2p+1}^{k,(k+2)/4,k}$ and the character is given by
eq. (19) with $z=1$.

5) $k=1\mod 4$, with $\twidle p=2\mod4$ we have:
$M\approx E_7^{k,(k+3)/4,k}$ with the character given
by eq. (19) with $z=1$.

6) $k=-1\mod 4$, $\twidle p=2\mod4$: $M\approx 
A_3^{1,(k+5)/4,2}$. The characters are given by eq. (19)
with $z=(k-1)/2$.
 
This concludes all the allowed cases.

Let us discuss now the extended algebra. In some cases
we have, in fact, a realization in terms of affine algebra.
For example, at $k=2$ all the bosonic lattices are given
for all $q$ by $D_n$, for some $n$. The full theory is
then $SO(2n+1)_1$. The decomposition eq. (3) is the usual
decomposition for $SO(2n+1)$ into  $n$ bosons 
and a fermion.
 
Now, in all the cases, we have an extended current
$$X(z)=\psi_1(z) \n e^{i\alpha \phi},\e$$
where $\psi_1(z)$ is the first parafermion of dimensions
$\Delta=1-1/k$. Thus the dimension of this current is
$$\Delta_X=1-1/k+\alpha^2/2,\e$$
which is always an integer. It is straight forwards to
compute $\Delta_X$ from the list above, and we find,
typically, that $\Delta_X=\twidle p+1$, where 
$\twidle p=0,1,2,3\ldots $. In the OPE of $X$ with $X^\dagger$
($X$ with $X$ is non--singular) we find only $T$
(the stress tensor)
and $J$ ($U(1)$ current). Thus we denote by $A_p$ the algebra
generated by $X$, $X^\dagger$, $T$ and $J$ (along with the 
derivatives and products,
of course, of $T$ and $J$). $A_0$ is the usual $SU(2)$
affine algebra. $A_{1/2}$ (by continuation: the algebras
and the models can be defined for $p$ half intgral,
although the fusion rules no longer coincides with $SU(2)$
affine ones; the algebra is then $A_p$ with half integral
$p$) is the $N=2$
superconformal algebra. The rest of the algebras appear to be
new.

The blocks of these theories, with respect to the algebra
$A_p$ are in 1--1 relation with ordinary $SU(2)$ obeying
also the same fusion rules. The modular matrix is a 
permutation of the $SU(2)$ one.
 
The exact OPE of $X$ with $X^\dagger$ is easily determined. 
For example, for $p=1$ we find
$$\twoline{
X(z) X^\dagger(u)={1\over (z-u)^4}+
{J(u)\over (z-u)^3}+{aT_{p}(u)+\half \partial J+\half JJ
\over (z-u)^2}+
}{
{2a\partial T_p+aJT_p+{1\over6}JJJ+\half J\partial J+{1\over6}\partial^3 J
\over (z-u)}+{\rm regular},}$$
where $J(u)=i\alpha\partial\phi$ is the $U(1)$ current,
$a=2\Delta(\psi_1)/c_{p.f.}$, $T_p=T+\half (\partial\phi)^2$ is the stress tensor minus the $U(1)$ part.
The rest of the OPE's ($X$ and $X^\dagger$ with $J$ and $T$) are 
obvious and we omit them for brevity, as $X$ is a primary
field with the $U(1)$ charge $\alpha^2$. 
The commutation relations for the moments of the algebra can
be easily written by standard methods \r\BPZ.
It is important
that the algebra $A_2$ (as the others) is expressed solely
in terms of $X$, $X^\dagger$, $T$ and $J$, as this
guarantees its universality. The parameters above depend only
on $p$ and the central charge. 
The representations of the algebra $A_p$ have two quantum numbers $\Delta$ 
(the dimension of the field), and 
$q$ (the $U(1)$ charge).

Let us digress now to describe the minimal models of the algebra $A_p$.
These are labeled, for each $p$, by an integer $k$. The central charge is
$$c={3k\over k+2},\e$$
i.e., the same as $SU(2)$ ($p=0$ case). We add one free bosons to the parafermions,
$$X=\psi^0_2 \n e^{i\alpha\phi}\nn,\e$$
where $\alpha=\sqrt{2(kp+1)/k}$.
Now, acting $k$ times with $X$, the parafermion drops and we get a net 
shift in momenta of $k\alpha=\sqrt{2(kp+1)k}$. This means that the characters
are expressed in terms of a theta function at the level $k(kp+1)$. Further,
acting with a single $X$ gives a shift of $2$ in the parafermion and a shift
of $\alpha$ in the bosonic momenta. Thus, the characters are given by
$$\chi^l_{m,\pr m}=\sum_{r\mod k} c^l_{m+2r} \Theta_{\pr m+2(kp+1)r;k(kp+1)}.
\e$$
This equation holds for $p$ integer or half integer. Now, the primary fields
are obtained by the demand that $X_n$ and $X^\dagger_n$ vanish for negative $n$.
For integer $p$, we find
$${|m-\pr m|\over k}\leq p,\e$$
and is integral.
For $p$ half inetgral in the $NS$ sector it is the same, and in the $R$ 
sector, it is half integral. 
We believe that these are the minimal models for the algebras
$A_p$ with $p=0,1/2,1,3/2,2,\ldots$.

For algebras, other than $SU(2)$, we similarly get extended
algebras for the realizations. The calculation of the OPE's is more difficult, but straightforwrd.

The algebras $A_p$ actually can be extended to two parameters, $\alpha^2$
and $c$, the central charge. More general minimal models can be easily
obtained by taking $X=\psi^0_m\exp(i\alpha\phi)$ for some general $m$.
The discussion above, for $m=2$, readily generalizes to any $m$.\foot{I
thank B. Noyvert for a discussion of this point, as well as pointing out
some typos.}

We can actually use the duality relation,
$$SU(n)_m\times SU(m)_n \times U(1)\approx U(nm)_1,\e$$
to express the theories $SU(n)_q$ with $q=-1+(k+g)\twidle q$,
in terms of $SU(m)_n$ parafermions and 
 bosons. This proceeds
by first writing
$$SU(n)_{-m/q}=SU(m)_{n/q}\times{\rm free\ bosons},\e$$
and then proceeding as before. It can be seen that the equation for the
dimension of the field, eq. (6), work as before. We omit the detail.

Finally, the reader must be curious about the general pseudo-affine theory. How
is it realized? We believe the answer lie in multi-parafermion theories.
It was noted already by Zamolodchikov and Fateev \r\Para\ that the monodromy of
parafermions allows for a more general solution,
$$\Delta(\psi_n)=q n(k-n)/k+M_n,\e$$
(for $SU(2)$; for other algebras it is similar), where $\psi_n$ is the $n$th
$Z_k$ parafermion, and $M_n$ are some integers. The multi-parafermions are for
the large part, yet to be explored. However, we believe that the central
charge is
$$c(q)=q c(1) \mod 4,\e$$
and thus by gluing them with free bosons it is possible to get the central
charge of the pseudo affine systems. We hope to report on this in the future.

\ACK
It is a pleasure to thank E. Baver for his participation in the early stages
of this work.

\refout
\bye